\documentclass[referee]{raa}            

\usepackage{graphicx} 
\usepackage{natbib}
\usepackage{amssymb, amsmath}
\bibpunct{(}{)}{;}{a}{}{,}

\usepackage[pagebackref=true]{hyperref}

\begin{document}

\title{Balmer Decrement and IRX Break in Tracing Dust Attenuation at Scales of Individual Star-forming Regions in NGC\,628}

   \volnopage{Vol.0 (20xx) No.0, 000--000}      
   \setcounter{page}{1}          

   \author{Man~Qiao 
      \inst{1,2}
   \and Mingfeng~Liu 
      \inst{3}
   \and Zongfei~Lyu 
      \inst{1,2}
   \and Shuang~Liu 
      \inst{1,2}
   \and Chao~Yang 
      \inst{1,2}
   \and Dong~Dong~Shi
      \inst{4}
   \and Fangxia~An
      \inst{1}
   \and Zhizheng~Pan
      \inst{1}
   \and Wenhao~Liu
      \inst{1}
   \and Binyang~Liu
      \inst{1}
   \and Run~Wen
      \inst{1,2}
   \and Yu~Heng~Zhang
      \inst{1,2}
   \and Xian~Zhong~Zheng 
      \inst{5}
   }

   \institute{Purple Mountain Observatory, Chinese Academy of Sciences, 10 Yuanhua Road, Nanjing 210023, China\\
      \and
         School of Astronomy and Space Science, University of Science and Technology of China, Hefei 230026, China\\
      \and
         Department of Physics and Institute of Theoretical Physics, Nanjing Normal University, Nanjing 210023, China\\
      \and
         Center for Fundamental Physics, School of Mechanics and Optoelectronic Physics, Anhui University of Science and Technology, Huainan 232001, China\\
      \and
         Tsung-Dao Lee Institute and Key Laboratory for Particle Physics, Astrophysics and Cosmology, Ministry of Education, Shanghai Jiao Tong University, Shanghai 201210, China; {\it xzzheng@sjtu.edu.cn}\\
\vs\no
   {\small Received 20xx month day; accepted 20xx month day}}

\abstract{ 
We investigate the relationships between infrared excess (IRX=$L_{\rm IR}/L_{\rm UV}$) and Balmer decrement (${\rm H}\alpha/{\rm H}\beta$) as  indicators of dust attenuation for 609 H\,{\small II} regions at scales of $\sim 50-200$\,pc  in NGC\,628, utilizing data from AstroSat, James Webb Space Telescope (JWST) and Multi Unit Spectroscopic Explorer (MUSE). Our findings indicate that about three fifths of the sample H\,{\small II} regions reside within the regime occupied by local star-forming galaxies (SFGs) along the dust attenuation correlation described by their corresponding color excess parameters $E(B-V)_{\rm IRX} = 0.51\,E(B-V)_{{\rm H}\alpha/{\rm H}\beta}$. Nearly 27\% of the sample exhibits $E(B-V)_{\rm IRX}> E(B-V)_{{\rm H}\alpha/{\rm H}\beta}$, while a small fraction ($\sim 13\%$) displays significantly lower $E(B-V)_{\rm IRX}$ compared to $E(B-V)_{{\rm H}\alpha/{\rm H}\beta}$.  These results suggest that the correlation between the two dust attenuation indicators no longer holds for spatially resolved H\,{\small II} regions.  
Furthermore, the ratio of  $E(B-V)_{\rm IRX}$ to $E(B-V)_{{\rm H}\alpha/{\rm H}\beta}$ remains unaffected by various physical parameters of the H\,{\small II} regions, including star formation rate (SFR), SFR surface density, infrared luminosity ($L_{\rm IR}$), $L_{\rm IR}$ surface density, stellar mass, gas-phase metallicity, circularized radius, and the distance to galactic center. We argue that the ratio is primarily influenced by the evolution of surrounding interstellar medium (ISM) of the star-forming regions, transitioning from an early dense and thick phase to the late blown-away stage. 
}

   \authorrunning{M. Qiao et al}            
   \titlerunning{Spatially Resolved ${\rm H}\alpha/{\rm H}\beta$ and IRX  in NGC\,628}  

   \maketitle

%
%
\section{Introduction}           
\label{sect:intro}

The interplay between stars, gas, metals and dust within galaxies involves complex processes that govern star formation history, chemical enrichment and structural growth. These interactions ultimately influence the appearance of galaxies and their spectral energy distribution (SED). Among these processes, dust attenuation plays a crucial role in reshaping observable characteristics of galaxies \citep{Draine2003, Galliano2018, Salim2020}.  Although dust comprises only a small fraction ($\sim$1\%) of the interstellar medium (ISM) in star-forming galaxies (SFGs), but the thermal radiation emitted in the far-infrared  --- resulting from dust absorbing the ultraviolet (UV) and optical radiation primarily from young stars ---  accounts for nearly half of the total energy budget of the extragalactic radiation background \citep[e.g.][]{Dole2006}.   Therefore, a better understanding of dust attenuation is essential for accurate measurements of galaxy observables \citep{Kennicutt1998, Calzetti2000, Buat2005} and for a comprehensive understanding of the cosmic baryon cycle \citep{Peroux2020}. 

Galaxy dust attenuation quantifies the fraction of light that is diminished by dust through absorption and net scattering (both away from and back into the line of sight), as well as the effects arising from the star-dust geometry \citep{Salim2020}. In practice, the slope of the ultraviolet (UV) continuum (1216 -- 3000\AA, noted as $\beta$), the Balmer decrement (${ \rm H}\alpha/{\rm H}\beta$), and the infrared excess (IRX: the ratio of infrared to UV luminosity) are commonly used as indicators of dust attenuation.  Balmer lines originate from H\,{\small II} regions, which typically have lifetimes of approximately $10^7$ years and are often surrounded by dense molecular clouds located in the inner disk component of a galaxy \citep{Kennicutt2012}. Therefore, the Balmer decrement records a stronger dust attenuation compared to the average across the entire galaxy.  The infrared excess (IRX) associates the infrared (IR) emission with the UV and optical radiation absorbed and re-emitted by dust, making it a good proxy for the average dust column density in SFGs where UV-emitting stars and dust are homogeneously mixed \citep{Qin2024}. IRX accounts for the global dust attenuation not only in the H\,{\small II} regions but also in the diffuse ISM. The UV slope $\beta$ reflects the reddening of the intrinsic stellar UV continuum, influenced by dust opacity, the dust attenuation curve, and the intrinsic slope \citep{Qin2022}.

The relationships among IRX, ${\rm H}\alpha/{\rm H}\beta$, and $\beta$ have been extensively examined in the literature. However, the results have often been controversial.  The IRX--$\beta$ relation demonstrates a positive correlation between the two indicators \citep{Meurer1995, Meurer1999, Liang2021, Duffy2023, Hamed2023}, but the scatter around this correlation is suggested to be driven by variations in the slope of the dust attenuation curve, the ratio of the present to past averaged star formation rate (SFR)  \citep{Kong2004, Burgarella2005}, and the star-dust geometry \citep{Wang2018}.  
In comparison to IRX, ${\rm H}\alpha/{\rm H}\beta$ tends to overestimate overall dust attenuation in galaxies. This has led to  a ratio of $E(B-V)$ (color excess, defined as the attenuation in the $B$-band minus that of in the $V$-band) between IRX and ${\rm H}\alpha/{\rm H}\beta$ being reported as 0.44 in some studies \citep{Calzetti2001, Wuyts2013}, in contrast to a median value of 0.51 for the ratio that varies with the specific SFR surface density \citep{Qin2019b}.  Given that IRX is found to be  jointly determined by SFR, galaxy size ($R_{\rm e}$), axial ratio ($b/a$) and metallicity but stellar mass ($M_\ast$), following a universal IRX scaling relation \citep{Qin2019a, Qiao2024, Qin2024}, the correlations between one dust attenuation indicator and other galaxy parameters (e.g. ${\rm H}\alpha/{\rm H}\beta$ versus $M_\ast$; \citealt{Zahid2017}) are likely affected by their dependence on these four parameters (i.e., SFR, $R_{\rm e}$, $b/a$ and metallicity), and may thus manifest as indirect and non-causal correlations.  
Furthermore, previous studies have mostly treated galaxies as unresolved systems, examining the correlations between dust attenuation indicators in a statistical manner. A spatially resolved investigation into how radiation from newly-formed stars is attenuated by the surrounding dust in both the dense and diffuse ISM components, as well as how contributions from different components contribute to overall attenuation, will provide insights into reconciling the differing results.

With high-resolution imaging observations in the optical with the \textit{Hubble Space Telescope} (HST) and in the near-IR and middle-IR with the \textit{James Webb Space Telescope} (JWST),   the stellar and dust emission in nearby galaxies can be spatially resolved down to the scales of individual H\,{\small II} regions. \citet{Kruijssen2019} unraveled that in nearby SFGs Balmer line emission from the H\,{\small II} regions and CO emission from molecular clouds no longer trace each other at spatial scales of $<\sim$100\,pc. In this work, we utilize the high-resolution multi-wavelength data of NGC\,628 (i.e. M74) to carry out a spatially resolved study of IRX and ${\rm H}\alpha/{\rm H}\beta$ to scales of H\,{\small II} regions, aimed at investigating the differences between  the two in tracing dust attenuation. 

This paper is organized as follows: In Section~\ref{sect:data}, we present a brief description of the data. Section~\ref{sect:sample} provides data reduction and analysis. Our results are shown in Section~\ref{sect:results}. We discuss our results in Section~\ref{sect:discussion} and give a summary in Section~\ref{sect:summary}. A standard $\Lambda$CDM cosmology with $H_0 = 70\,{\rm km^{-1}\,Mpc^{-1}}$, $\Omega_\Lambda = 0.7$ and $\Omega_{\rm m} = 0.3$, and a \citet{Chabrier2003} Initial Mass Function (IMF) are adopted throughout the paper.

\begin{table}
   \begin{center}
   \caption[]{Basic Parameters of NGC\,628.}\label{Tab1}
   \begin{tabular}{clcl}
   \hline\noalign{\smallskip}
   Parameter & Value & Reference \\
   \hline\noalign{\smallskip}
   R.A. (J2000) & $01^{h}36^{m}41.^{s}747$ & NASA/IPAC Extragalactic Database (NED) \\
   Dec. (J2000)  & +$15^{d}47^{m}01.^{s}18$ & NASA/IPAC Extragalactic Database (NED) \\
   Distance & 9.84\,Mpc & \citet{Anand2021} \\
   Stellar Mass & $10^{10.24}\,{\rm M_\odot}$ & \citet{Leroy2021} \\
   Star Formation Rate & 1.73 ${\rm M_\odot\,yr^{-1}}$ & \citet{Leroy2021} \\
   Inclination & $8.9^\circ$ & \citet{Lang2020} \\
   Position Angle & $20.7^\circ$ & \citet{Leroy2021} \\
   Morphological Type & SA(s)c & \citet{Buta2015} \\ 
   \noalign{\smallskip}\hline
   \end{tabular}
   \end{center}
\end{table}

\section{Data}
\label{sect:data}

NGC\,628 is a star-forming spiral galaxy nearly face-on and its basic parameters are given in Table~\ref{Tab1}.
NGC\,628 has been extensively observed and multi-band high-resolution imaging and spectroscopic data are publicly available. We make use of spectroscopic data from the Physics at High Angular resolution in Nearby GalaxieS (PHANGS)-Multi Unit Spectroscopic Explorer (MUSE) survey \citep{Emsellem2022}, the IR data from the PHANGS-James Webb Space Telescope (JWST) survey \citep[project-ID 02107;][]{Lee2023}, and the UV data from the PHANGS-AstroSat survey \citep{Hassani2024}.

\subsection{PHANGS-MUSE}

PHANGS-MUSE made use of the MUSE integral field spectrograph on board the Very Large Telescope (VLT) at European Southern Observatory (ESO) mapping 19 massive ($9.4 < \log\,M_\ast/{\rm M_\odot} < 11.0$) nearby ($ D \lesssim 20$\,Mpc) star-forming disk galaxies \citep{Emsellem2022}. The science-ready data cubes and data products are available \footnote{\url{https://www.canfar.net/storage/vault/list/phangs/RELEASES/PHANGS-MUSE}} (DR1.0). \citet{Congiu2023} presented a catalogue of ionized nebulae distributed across the 19 galaxies of the PHANGS-MUSE sample, using a new model-comparison-based algorithm that exploited the principle of the odds ratio to assign a probabilistic classification to each nebula. The catalogue contains spectral and spatial information for over 40\,000 ionized nebulae.  In our study, we utilize this ionized nebula catalogue to select H\,{\small II} regions within NGC\,628, providing positions, radius, ${\rm H}\alpha$ flux and ${\rm H}\beta$ flux of each of the selected H\,{\small II} regions of our sample (see Section~\ref{sec:HIIsam} for more details).

\subsection{PHANGS-JWST}

PHANGS-JWST is a JWST Cycle~1 treasury program to image 19 nearby ($D < 20$\,Mpc) main-sequence spiral galaxies at spatial scales of $\sim 5-50$\,pc in eight JWST filters from 2 to 21\,$\mu$m using the Near Infrared Camera (NIRCam) through filters F200W, F300M, F335M, and F360M, and the Mid-InfraRed Instrument (MIRI) through filters F770W, F1000W, F1130W, and F2100W \citep{Lee2023, Williams2024}. The unprecedented sensitivity and angular resolution of JWST enable us to map infrared emission from the young stellar populations and dusty ISM across the entire disks of sample galaxies. \citet{Williams2024} presented a full public data release\footnote{\url{https://archive.stsci.edu/hlsp/phangs/phangs-jwst}} (DR1.0.1) from the PHANGS-JWST Cycle~1 treasury program. NGC\,628 is one of the 19 galaxies, with JWST eight-band imaging observations covering the central region of $3\farcm8 \times 2\farcm2$ (11\,kpc $\times$ 6.3\,kpc) \citep{Watkins2023}. The pixel scales of these images are $0\farcs03$ (F200W), $0\farcs06$ (F300M, F335M, F360M), $0\farcs11$ (F770W, F1000W, F1130W, F2100W). We use these imaging data of NGC\,628 to derive IR luminosity ($L_{\rm IR}$) of the selected H\,{\small II} regions in our sample.

\subsection{PHANGS-AstroSat}

PHANGS-AstroSat contains  multi-band UV imaging of 31 massive nearby ($D<22$\,Mpc) spiral galaxies observed with the Ultraviolet Imaging Telescope \citep[UVIT;][]{Kumar2012, Tandon2017} on the AstroSat satellite \citep{Singh2014} in the  wavelength range of $1480-2790$\,\AA\ \citep{Hassani2024}. The adopted UV filters include one narrowband filter (N279N), three broadband filters (F148W, F154W, and N242W), and five medium-band filters (F169M, F172M, N219M, N245M, and N263M). These UV images have a pixel scale is $0\farcs4176$ and the median angular resolution of $1\farcs4$ (corresponding to a physical scale between 25 and 160\,pc). We use the background-subtracted, foreground-corrected images of NGC\,628 from CANFAR\footnote{\url{https://www.canfar.net/storage/vault/list/phangs/RELEASES/PHANGS-AstroSat}}, and calculate the UV luminosity ($L_{\rm UV}$) for each of the H\,{\small II} regions in our sample.

\section{Data Reduction and Analysis}
\label{sect:sample}

The UV, optical and IR science images of NGC\,628 are obtained with different observing facilities and have distinct Point Spread Functions (PSF). We derive the corresponding PSF for each band image and perform image convolution to match all images to the same PSF. The same aperture is applied to all PSF-matched images in order to derive the multi-band photometry for given targets. 

\subsection{PSF Construction}

We extract empirical PSF using point sources in each of the AstroSat/UVIT science images. Firstly, we identify the point sources in the background-subtracted and foreground-corrected images of NGC\,628 by crossmatching the detected objects with the star catalog from Gaia Data Release 3 \citep[DR3;][]{Gaia2023}, and co-add the UV images of these stars to obtain a stacked star image. Secondly, we fit a two dimensional Moffat PSF model \citep{Moffat1969} to the stacked star image, and adopt the best-fitting model to be the empirical PSF for the given-band UV image.  This best-fitting PSF is characterized by a compact core and extended wings \citep{Leahy2020, Hassani2024}. Doing so,  we obtained empirical PSFs for all UV science images, giving the full-width at half maximum (FWHM) to be $1\farcs393$, $1\farcs522$, $1\farcs218$, $1\farcs451$, $1\farcs644$, $1\farcs212$, $1\farcs104$, $1\farcs015$, and $1\farcs017$ for PSFs in F148W, F154W, F169M, F172M, N219M, N242W, N245M, N263M, and N279N, respectively.

The actual JWST NIRCam and MIRI science images of NGC\,628 are fully filled by the target galaxy and it is hard to find a sample of isolated stars for constructing the empirical PSFs  \citep{Hoyer2023}.  We use the software tool WebbPSF\footnote{\url{https://webbpsf.readthedocs.io/en/stable/index.html}} \citep{Perrin2012, Perrin2014} to generate PSFs for given JWST images.  Moreover, JWST's PSF varies in both the spatial and temporal dimensions. WebbPSF is suggested to be a better way to precisely generate PSFs for the JWST NIRCam and MIRI images \citep{Nardiello2022}. 

\subsection{PSF Matching}

We aim to derive dust attenuation from the UV and IR fluxes of spatially resolved H\,{\small II} regions within NGC\,628. It is necessary to match all images to the same PSF. This can be done by resampling high-resolution images to have the same pixel scale as the image of the lowest resolution and then convolve the resampled image with corresponding convolution kernels. Among the science images of all bands that we use, the UV images from AstroSat/UVIT have the lowest resolution, and UVIT-N219M image has the largest PSF FWHM. We thus resample JWST images to have the same pixel scale ($0\farcs4176$) and area coverage as the AstroSat/UVIT images. Then, JWST and AstroSat/UVIT PSFs are used to generate convolution kernels using the convolution kernels generated via the Python-based PSF Homogenization kERnels (PyPHER)\footnote{\url{https://pypher.readthedocs.io/en/latest/}} \citep{Boucaud2016}.  The kernels are used to convolve the corresponding eight resampled JWST and eight UVIT images, and obtained the PSF-matched JWST and UVIT images to the AstroSat/UVIT-N219M images. 

\begin{figure*} 
   \begin{center}
   \includegraphics[width=0.8\textwidth]{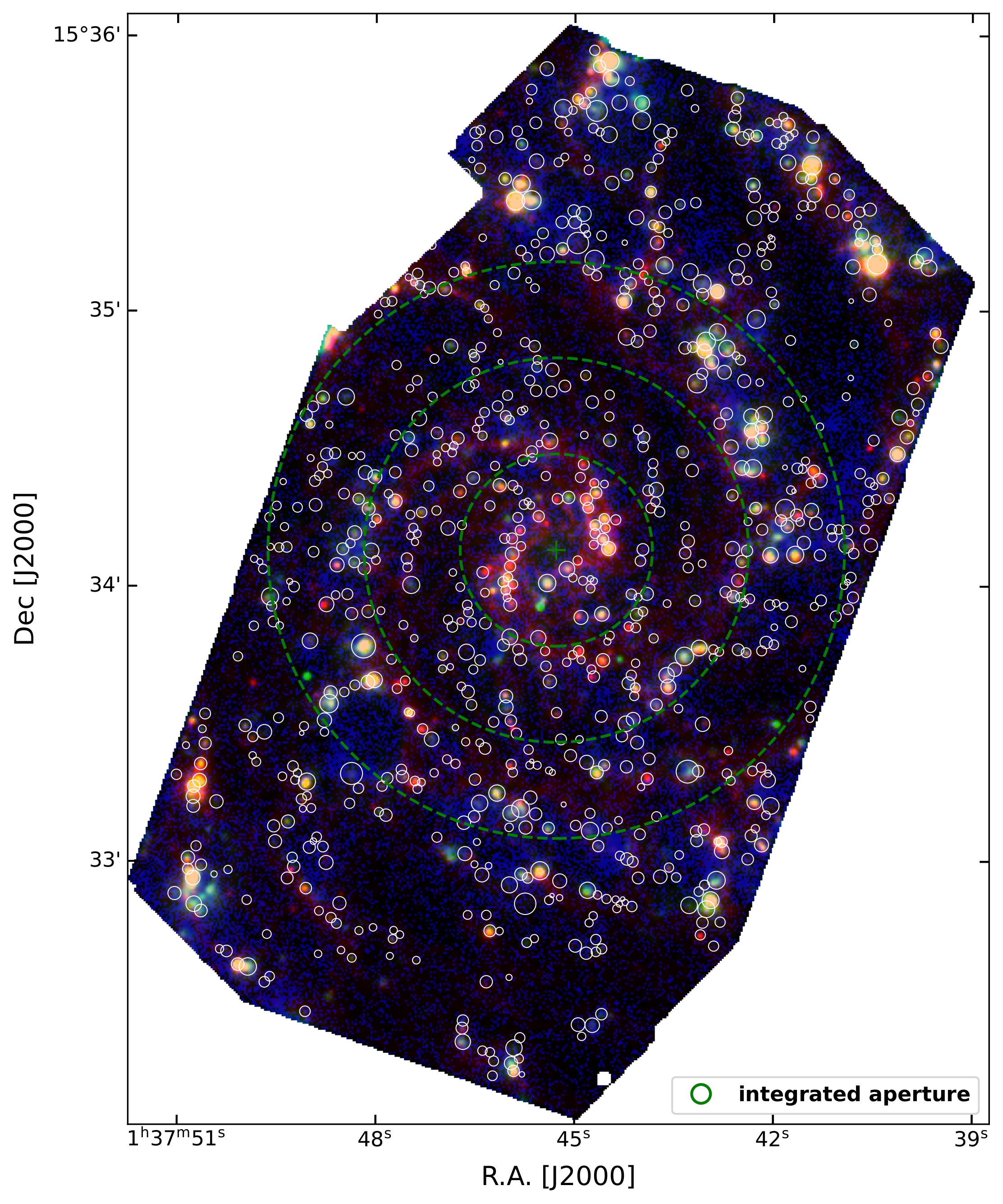}
   \caption{Distribution of 609 H\,{\small II} regions for NGC\,628. The white solid circles represent the photometric apertures of the final 609 H\,{\small II} regions of NGC\,628 in our sample with circularized radius, and the background refers to the RGB image of NGC\,628 obtained by combining the JWST MIRI F2100W infrared image (red), the MUSE H$\alpha$ map (green), and AstroSat UVIT N148W ultraviolet image (blue). The green dashed circles demonstrate the three apertures we used to analyze the integrated dust attenuation, with radii of 1.0, 2.0, and 3.0\,kpc.}
   \label{Fig1} 
   \end{center}
\end{figure*}

Additionally, to create the RGB image of NGC\,628, we extracte the H$\alpha$ map of NGC\,628 from the MUSE data cubes, and resample the MUSE H$\alpha$ map to the same pixel scale as the AstroSat UVIT images ($0\farcs4176$). Figure~\ref{Fig1} presents the RGB image of NGC\,628, where the red, green and blue channels correspond to the JWST/MIRI F2100W image, MUSE H$\alpha$ map and AstroSat/UVIT F148W image, respectively.

\subsection{Selection for H\,{\small II} Regions}\label{sec:HIIsam}

Our sample of  H\,{\small II} regions of NGC\,628 are selected from the ionized nebula catalogue given by \citet{Congiu2023}.  In total 40\,920 nebulae are detected in 19 nearby galaxies observed by the PHANGS-MUSE survey. These nebulae are identified from a channel-integrated flux image of [O\,{\small III}]$\lambda$5007, H$\alpha$ and [S\,{\small II}]$\lambda \lambda$6717,6731 lines, and classified into three types: H\,{\small II} regions, planetary nebulae and supernova remnants. In total 2\,798 H\,{\small II} regions are identified in NGC\,628. By limiting the H\,{\small II} regions in the overlap area of the MUSE, JWST, and AstroSat observations  \citep{Emsellem2022, Lee2023, Hassani2024}, we obtain a sample of 1\,921 H\,{\small II} regions in NGC\,628 with the MUSE, JWST and AstroSat observations after removing those at the edges of the JWST and AstroSat images. We use this sample of 1\,921 H\,{\small II} regions for our further analyses. 

\subsection{Aperture Photometry}

\begin{figure*}
   \centering
   \includegraphics[width=1.0\textwidth]{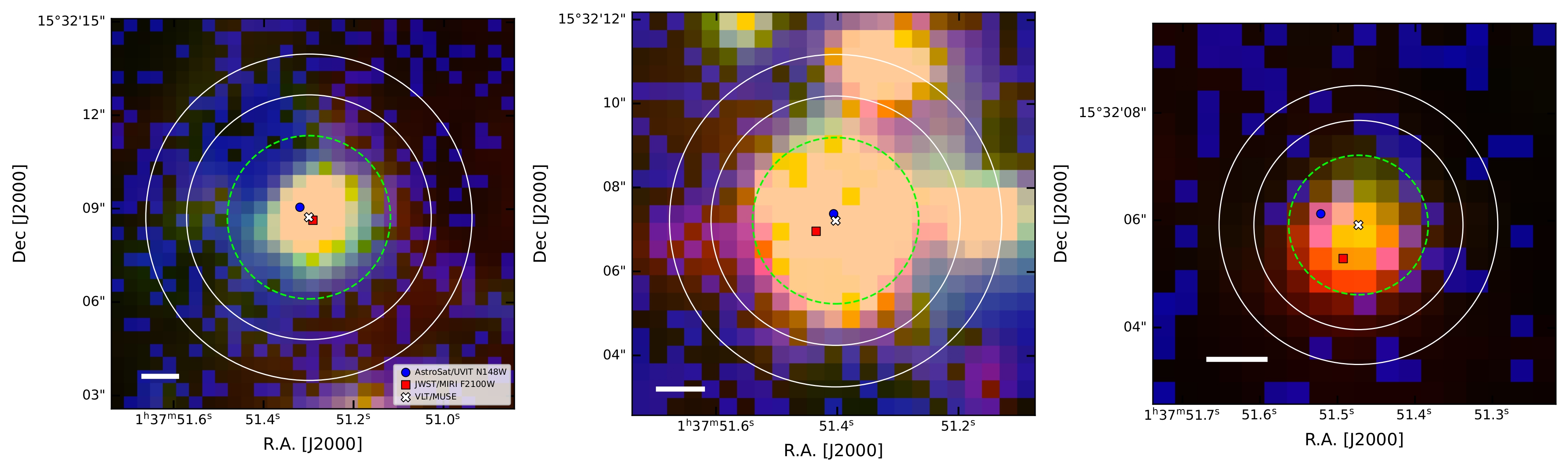}
   \caption{Example images of the selected aperture and annulus for three H\,{\small II} regions of different sizes, with radii of 125 pc, 94 pc, and 62 pc. The dashed green circles represent the photometric apertures with circularized radius of H\,{\small II} regions from the ionized nebula catalogue, while the solid white annuli are used to estimate the local background, with inner and outer radii of 1.5$r$ and 2$r$, where $r$ is the circularized radius of the target H\,{\small II} region. The RGB images are obtained by combining the JWST MIRI F2100W infrared image (red), the MUSE H$\alpha$ map (green), and AstroSat UVIT N148W ultraviolet image (blue). The white crosses, red squares and blue circles mark the centroid positions of H\,{\small II} regions in H$\alpha$, IR and UV bands. The solid white lines in lower-left corner indicates a spatial scale of 50\,pc.}
   \label{Fig2}
\end{figure*}

We notice that H\,{\small II} regions are located within the complex spiral structures of NGC\,628, making the photometric measurement of their UV and IR fluxes challenging. To reduce contamination from surrounding diffuse emissions,  we employ a circular annulus to estimate the local background, facilitating subtraction in aperture photometry of the targets. We use the software tool PHOTUTILS \citep{Bradley2024} to determine the local background and perform aperture photometry in the PSF-matched images from the AstroSat and JWST observations for the selected H\,{\small II} regions. The released JWST image sets include science and error images, while Astrosat image sets only provide science images.  We thus perform aperture photometry with errors derived from the error maps for JWST, and as 5\,percent for AstroSat following \citet{Hassani2024}.
When estimating the local background for each H\,{\small II} region, we first mask out surrounding H\,{\small II} regions to mitigate their contamination. We then calculate the local background using an annulus with inner and outer radii of 1.5$r$ and 2$r$, where $r$ is the circularized radius of the target H\,{\small II} region. By the way, we test varying annulus sizes for background estimation, and the results demonstrate that dust attenuation remains consistent. The aperture photometry with the same aperture sizes as ionized nebula catalogue given by \citet{Congiu2023} is performed on the UV and IR images. We provide example images of the selected aperture and annulus for three H\,{\small II} regions of different sizes, with radii of 125 pc, 94 pc, and 62 pc, as shown in Figure~\ref{Fig2}. The selected aperture is large enough to encompass the entire H\,{\small II} region, while the annulus is sufficiently distant to avoid contamination from the target.

\subsection{UV and IR Luminosities, and SFR}

\begin{figure*}
   \centering
   \includegraphics[width=0.8\textwidth]{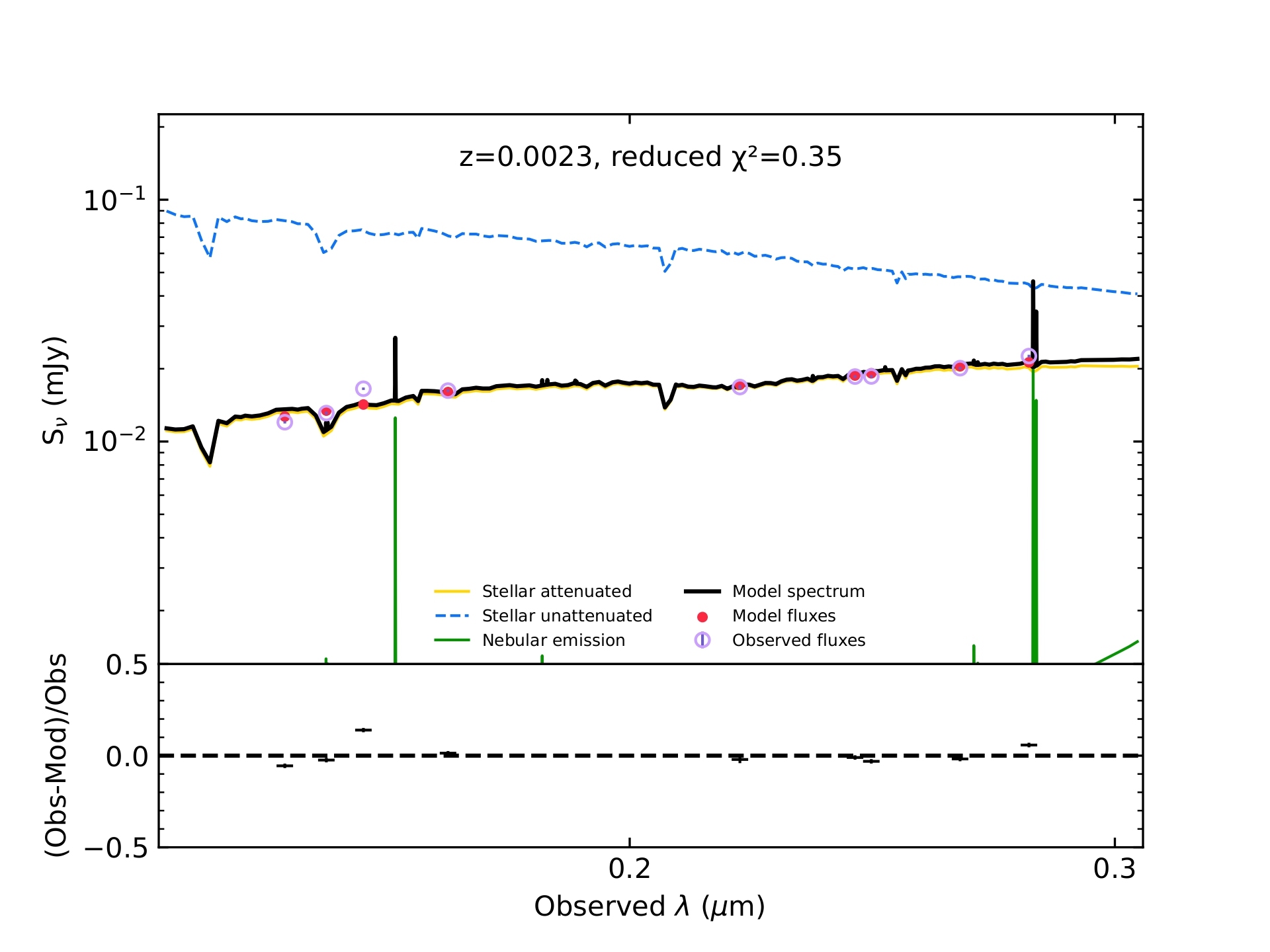}
   \caption{An example of the best-fitting UV spectrum model from CIGALE for the SED of an H\,{\small II} region.}
   \label{Fig3}
\end{figure*}

The UV and IR luminosities of  our sample H\,{\small II} regions are derived by fitting their SEDs composed of nine AstroSat UV bands and eight JWST NIR and mid-IR bands, using the software  Code Investigating GALaxy Emission \citep[CIGALE;][]{Boquien2019}. 
Stellar spectral templates with Chabrier stellar IMF \citep{Chabrier2003} and Solar metallicity from the \citet[][BC03]{Bruzual2003} stellar population synthesis model are adopted in the fitting. The star formation history is assumed to follow a delayed form: SFR$(t) \propto t/\tau^2\,e^{-t/\tau}$ for $0 \leq t \leq t_0$, where $t_0$ is the age of the onset of star formation, and $\tau$ is the time at which the SFR peaks. An additional recent decline starburst is also adopted, with the stellar age varying from 10 to 200\,Myr, and the e-folding time varying from 100 to 150\,Myr. 
The Calzetti attenuation law \citep{Calzetti2000} is adopted to count the dust attenuation with $E(B-V)$ ranging from 0.1 to 0.5\,mag.  Figure~\ref{Fig3} shows an example of the best-fitting model of the SED in the UV from CIGALE.

\begin{figure*}
   \centering
   \includegraphics[width=0.8\textwidth]{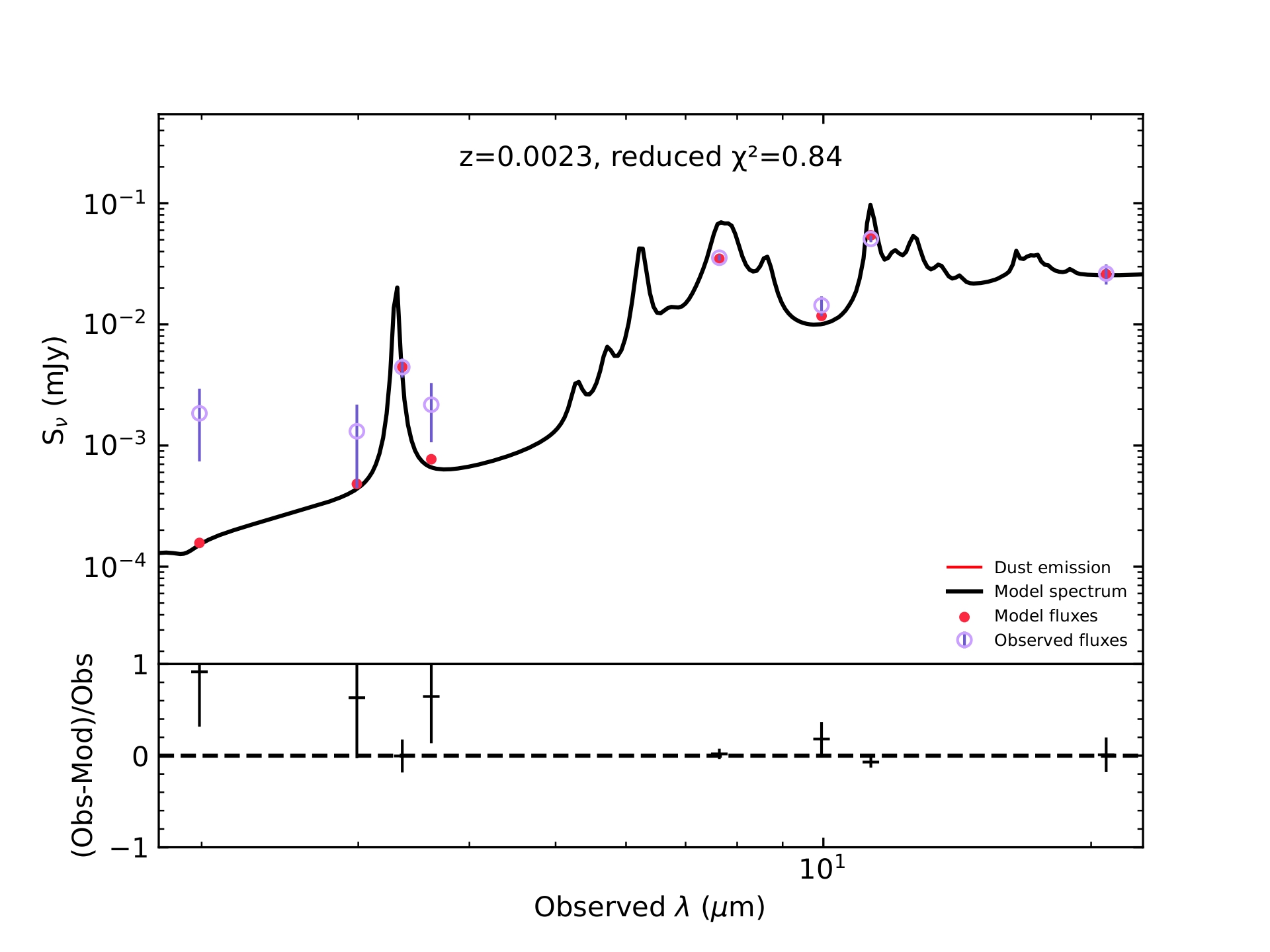}
   \caption{An example of the best-fitting IR spectrum model from CIGALE for the SED of an H\,{\small II} region.}
   \label{Fig4}
\end{figure*}

The dust emission SED templates from \citet{Draine2007} are employed in our SED fitting. We assume  mass fraction of polycyclic aromatic hydrocarbon (PAH) is in the range of [0.47, 4.58], and the fraction of illumination ranges over [0, 1]. The minimum and maximum radiation field are in the range of 0.1 to 25 and $10^3$ to $10^6$, respectively. Figure~\ref{Fig4} illustrates an example of the best-fitting  model to the observed dust SED.

The UV luminosity (1216$-$3000\,\AA) and the IR luminosity (8$-$1000\,$\mu$m) are calculated from the integration of the best-fitting SED. We also calculate the ${\rm SFR} = 1.09 \times 10^{-10}(L_{\rm IR} + 2.2 L_{\rm UV})$ \citep{Chabrier2003, Bell2005}, where $L_{\rm IR}$ and $L_{\rm UV}$ are given in units of $L_\odot$ with $L_\odot = 3.83 \times 10^{33}\,{\rm erg\,s^{-1}}$ and SFR is given in units of ${\rm {M_\odot\,yr^{-1}}}$ with ${\rm {M_\odot = 1.99 \times 10^{33}}}$\,g.

\subsection{Metallicity}

The strong-line ratios ([O\,{\small II}]\,$\lambda$3727+[O\,{\small III}]\,$\lambda \lambda$4959,\,5007)/H$_\beta$ (R23), [O\,{\small III}]\,$\lambda$5007/[O\,{\small II}]\,$\lambda$3727 (O32), [N\,{\small II}]\,$\lambda$6584/H$\alpha$ (N2), or ([O\,{\small III}]\,$\lambda$5007/H$_\beta$)/([N\,{\small II}]\,$\lambda$6584/H$\alpha$) (O3N2), are the most widely used metallicity indicators \citep[e.g.][]{McGaugh1991, Storchi-Bergmann1994, Kewley2002, Pettini2004, Maiolino2008, Zahid2014, Curti2017,Sanders2024}. To be consistent with \citet{Qin2019b}, we choose N2 to calculate the gas-phase metallicity of the H\,{\small II} regions in NGC\,628. The oxygen abundance (O/H) is estimated following the conversion formula given by \citet{Pettini2004} as ${\rm {12 + \log(O/H) = 0.32 \times N2^3 + 1.26 \times N2^2 + 2.03 \times N + 9.37}}$, with N2 = $\log$([N\,{\small II}]\,$\lambda$6584/H$\alpha$). The diffuse ionized gas (DIG)-corrected fluxes of [N\,{\small II}]\,$\lambda$6584 and ${\rm H}\alpha$ of our H\,{\small II} regions are from the ionized nebula catalogue. Note that the valid range of N2 is [$-$2.5, $-$0.3], corresponding  to ${\rm {12 + \log(O/H)}}$ in the range of [7.17, 8.86] \citep{Pettini2004}. Therefore, we exclude H\,{\small II} regions with N2 out of the valid range and keep the rest 1\,574 H\,{\small II} regions for our analyses.

\subsection{Dust Attenuation Parameters}

The IR excess parameter IRX can be directly obtained as  IRX = $L_{\rm IR}$/$L_{\rm UV}$ for the 1\,574   H\,{\small II} regions  in NGC\,628, and ${\rm H}\alpha/{\rm H}\beta$ come from the ionized nebula catalogue. The two dust attenuation indicators are then converted into color excess as $E(B-V)_{\rm IRX}$ and $E(B-V)_{{\rm H}\alpha/{\rm H}\beta}$, respectively.  The conversion is done following 
\begin{align}\label{eq:eq1}
   & {\rm E(B-V)}_{\rm IRX} = A_{\rm FUV}/k_{\rm FUV}, {\rm with} \nonumber \\
   & A_{\rm FUV} = -0.0333X^3 + 0.3522X^2 + 1.1960X + 0.4967\,\,{\rm and} \nonumber \\
   & X = \log(L_{\rm IR}/L_{\rm FUV}) = \log({\rm IRX}/1.38),
\end{align}
where $k_{\rm FUV} = 10.22$ for the adopted Calzetti attenuation curve \citep{Calzetti2000}, and the relation between $A_{\rm FUV}$ and IRX was given by \citet{Buat2005}. The 1.38 is the conversion factor from far-UV (FUV) luminosity ($L_{\rm FUV}$) to the UV luminosity, assuming a stellar population at age of 100\,Myr formed through a constant SFR history. And the $E(B-V)_{{\rm H}\alpha/{\rm H}\beta}$ is given by 
\begin{align}\label{eq:eq2}
   & E(B-V)_{{\rm H}\alpha/{\rm H}\beta}=A_{{\rm H}\alpha}/k_{{\rm H}\alpha}, {\rm with} \nonumber \\ 
   & A_{{\rm H}\alpha}=\frac{-2.5k_{{\rm H}\alpha}}{k_{{\rm H}\beta}-k_{{\rm H}\alpha}} \log \left( \frac{2.86}{{\rm H}\alpha/{\rm H}\beta} \right),
\end{align}
where $k_{{\rm H}\alpha}$ = 3.31 and $k_{{\rm H}\beta}$ = 4.60 for the adopted Calzetti attenuation curve at the wavelengths of the ${\rm H}\alpha$ and ${\rm H}\beta$ emission lines. And the factor of 2.86 is the intrinsic ${\rm H}\alpha/{\rm H}\beta$ line flux ratio, under the case B recombination condition with a temperature of $T = 10^4$\,K and an electron density of $10^2\,{\rm cm}^{-3}$ \citep{Osterbrock2006}. Furthermore, to quantify the color excess discrepancy between $E(B-V)_{\rm IRX}$ and $E(B-V)_{{\rm H}\alpha/{\rm H}\beta}$, the IRX-to-${\rm H}\alpha/{\rm H}\beta$ color excess ratio of $R_{\rm EBV} \equiv E(B-V)_{\rm IRX}/E(B-V)_{{\rm H}\alpha/{\rm H}\beta}$ is introduced as adopted in \citet{Qin2019b}. 

\section{Results}
\label{sect:results}

\begin{figure*} 
   \begin{center}
   \includegraphics[width=0.8\textwidth]{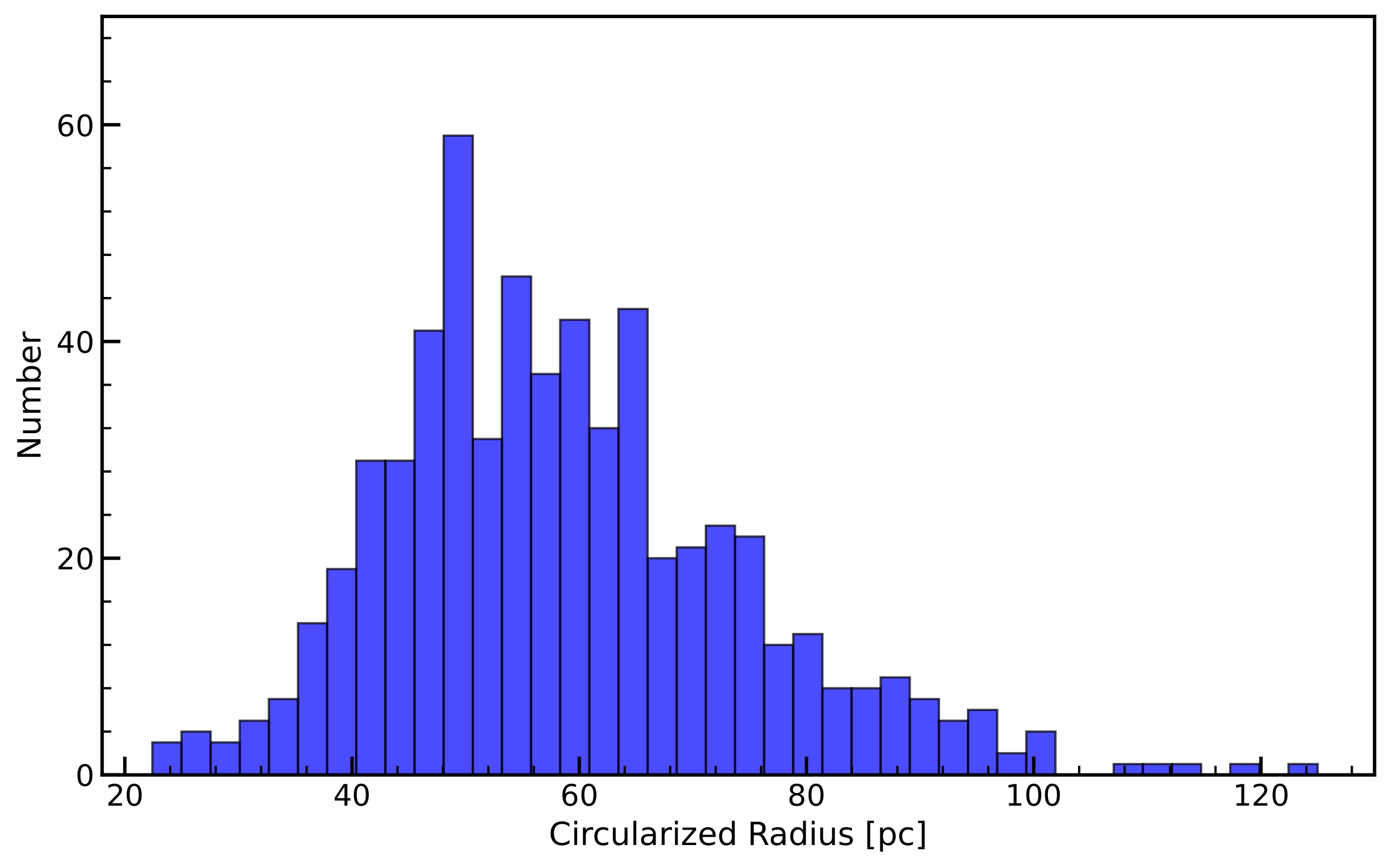}
   \caption{Histogram of circularized radius for 609 H\,{\small II} regions in NGC\,628.}
   \label{Fig5} 
   \end{center}
\end{figure*}

We want to examine connections of color excess between IRX and ${\rm H}\alpha/{\rm H}\beta$ at scales of H\,{\small iI} regions.  Due the photometric uncertainties, some measured fluxes are below zero. We get rid of the sample targets with negative values for fluxes of ${\rm H}\alpha$ or ${\rm H}\beta$ lines, UV or IR luminosities, and the estimated $E(B-V)_{{\rm H}\alpha/{\rm H}\beta}$ and $E(B-V)_{\rm IRX}$. The removed objects are faint in either Balmer-line emission or in UV and IR (dust) emission.  We obtain  609 data points having secure estimates of $E(B-V)_{{\rm H}\alpha/{\rm H}\beta}$ and $E(B-V)_{\rm IRX}$. Figure~\ref{Fig1} presents the distribution of these H\,{\small II} regions in NGC\,628. The 609 H\,{\small II} regions exhibit sizes (circularized radius) range from 22\,pc to 125\,pc, and the histogram of circularized radius for these H\,{\small II} regions is shown in Figure~\ref{Fig5}.

\begin{figure*}
   \centering
   \includegraphics[width=0.8\textwidth]{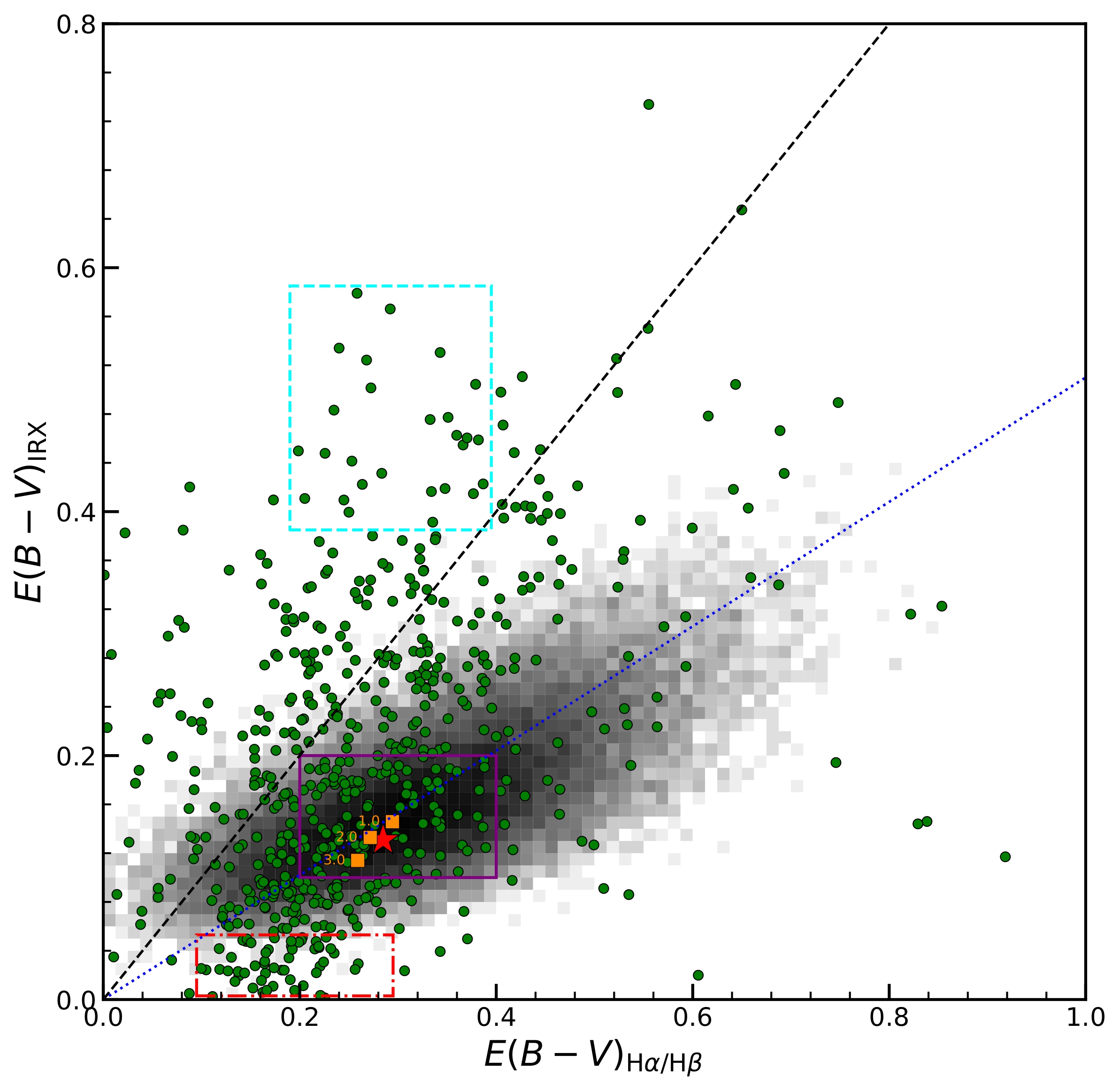}
   \caption{Comparison of $E(B-V)_{\rm IRX}$ and $E(B-V)_{{\rm H}\alpha/{\rm H}\beta}$ for 609 H\,{\small II} regions. The green points represent the 609 H\,{\small II} regions, and the background grey-scale map demonstrates the distribution of 32\,354 local SFGs from \citet{Qin2019b}. The black dashed line denotes the 1:1 line, and the blue dotted line shows the average ratio $E(B-V)_{\rm IRX}/E(B-V)_{{\rm H}\alpha/{\rm H}\beta} = 0.51$ given in \citet{Qin2019b}. The three orange squares represent the integrated $E(B-V)_{\rm IRX}$ and $E(B-V)_{{\rm H}\alpha/{\rm H}\beta}$ for H\,{\small II} regions within the three apertures shown in Figure~\ref{Fig1} and red star is for all H\,{\small II} region in our sample, respectively. And the labeled numbers denote the aperture size. The cyan (dashed), purple (solid), and red (dashed-dotted) rectangles arbitrarily mark the loci of three types of  H\,{\small II} regions in different evolutionary stages.}
   \label{Fig6}
\end{figure*}

Figure~\ref{Fig6} shows comparison of $E(B-V)_{\rm IRX}$ and $E(B-V)_{{\rm H}\alpha/{\rm H}\beta}$ for 609 H\,{\small II} regions in NGC\,628. One can see that the H\,{\small II} regions at scales of $\sim$200\,pc spread much wider than local SFGs. While the local SFGs (grey-scale map) apparently form a sequence in the sense that $E(B-V)_{\rm IRX}$ globally increases with $E(B-V)_{{\rm H}\alpha/{\rm H}\beta}$,  the majority of the sample H\,{\small II} regions fall into the locus area of the local SFGs,  suggesting that these H\,{\small II} regions are representative of these local SFGs in dust attenuation. However, some data points are located out the SFG area, e.g., these above the black dashed line or in the red box. We note that approximately 27\% of the sample H\,{\small II} regions distribute above the one-to-one line,  meaning that IRX probes a higher degree of obscuration than ${\rm H}\alpha/{\rm H}\beta$. Meanwhile, there is a portion of H\,{\small II} regions with relatively small $E(B-V)_{\rm IRX}$ and $E(B-V)_{{\rm H}\alpha/{\rm H}\beta}$, which are distributed in the lower-left corner of the $E(B-V)_{\rm IRX}-E(B-V)_{{\rm H}\alpha/{\rm H}\beta}$ diagram. We also calculate the integrated results of $E(B-V)_{\rm IRX}$ and $E(B-V)_{{\rm H}\alpha/{\rm H}\beta}$ by selecting apertures with three different radii, including 1, 2 and 3\,kpc, centered at the center of NGC\,628, as well as the results from the entire observed area. We find that the relationship between the integrated $E(B-V)_{\rm IRX}$ and $E(B-V)_{{\rm H}\alpha/{\rm H}\beta}$ agrees with that of the local SFGs, being near the median of $R_{\rm EBV} = 0.51$ given by \citet{Qin2019b}.

\begin{figure*}
   \centering
   \includegraphics[width=0.8\textwidth]{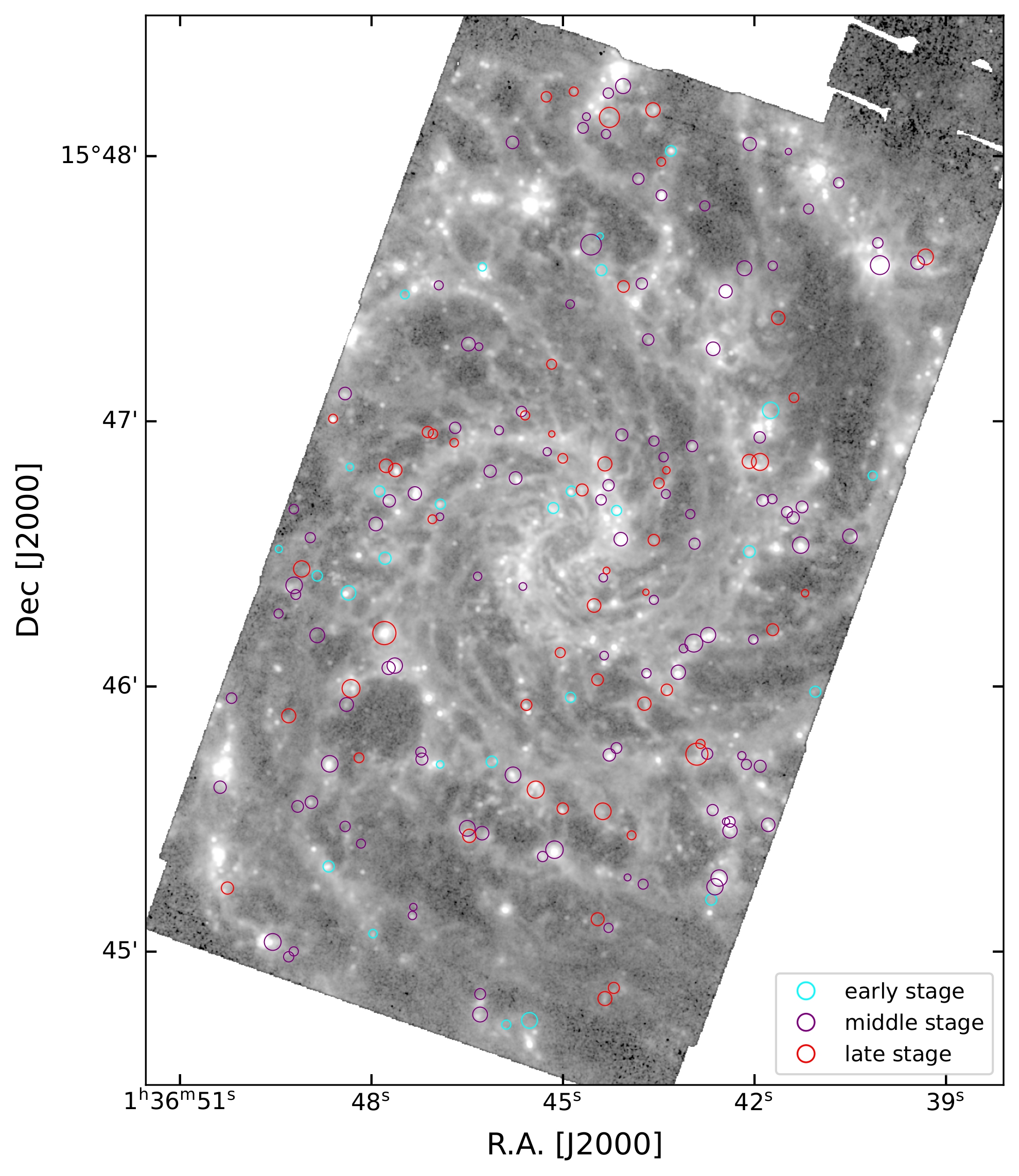}
   \caption{Locations of three types of H\,{\small II} regions enclosed in the three rectangles in Figure~\ref{Fig6}. The sizes of circles are proportional to the radii of given H\,{\small II} regions, shown in color same as the corresponding rectangles.}
   \label{Fig7}
\end{figure*}

\begin{figure*}
   \centering
   \includegraphics[width=0.8\textwidth]{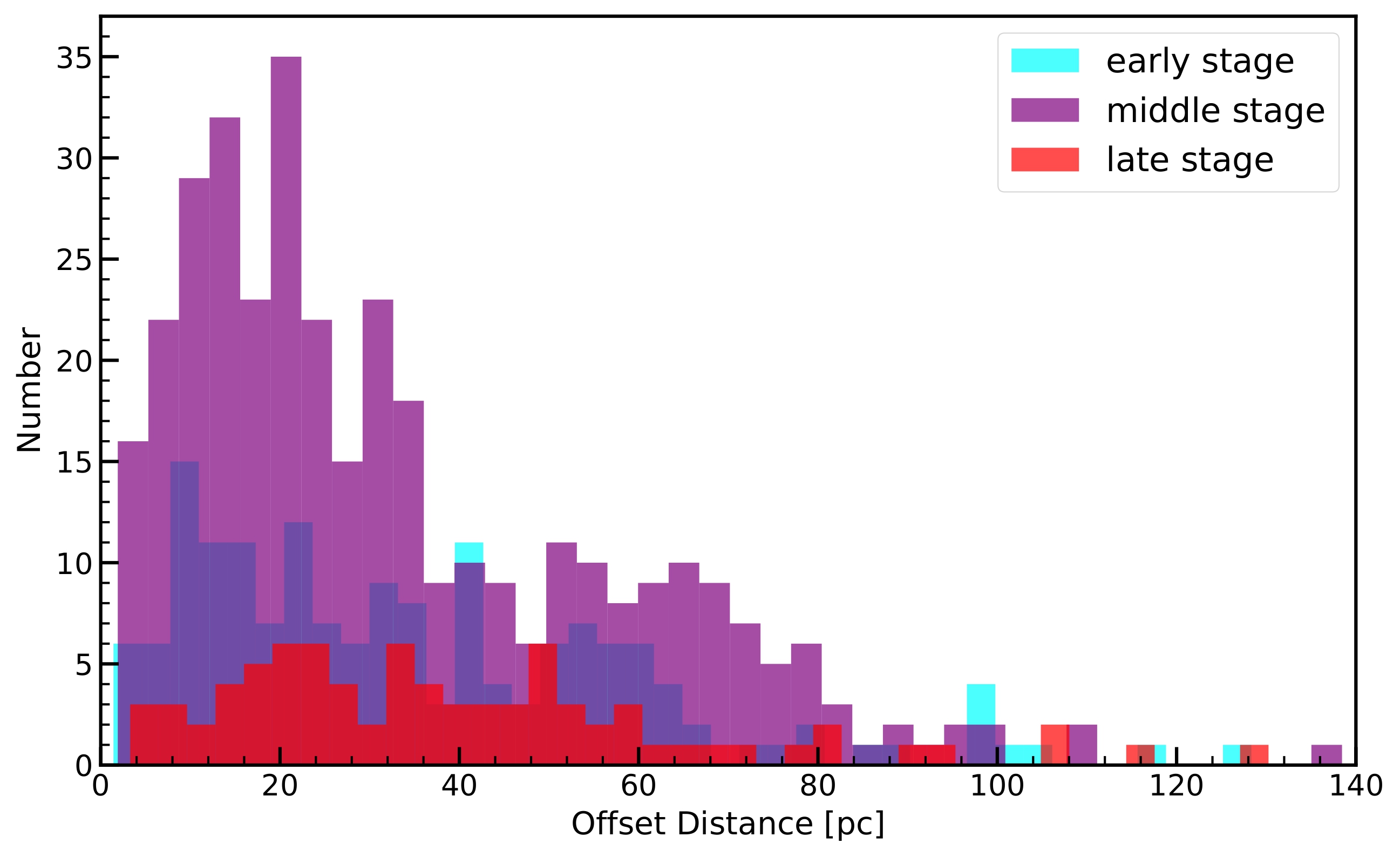}
   \caption{The histogram of central offsets between PHANGS-MUSE H$\alpha$ and PHANGS-JWST/MIRI 21\,$\mu$m for three types of H\,{\small II} regions. The colors of H\,{\small II} regions at different evolutionary stages are the same as Figure~\ref{Fig7}.}
   \label{Fig8}
\end{figure*}

We then investigate the locations of the selected H\,{\small II} regions in NGC\,628. We selected three groups of H\,{\small II} regions  being distinct in $R_{\rm EBV}$, i.e. located separated in the $E(B-V)_{\rm IRX}-E(B-V)_{{\rm H}\alpha/{\rm H}\beta}$ diagram as illustrated by the three rectangles in Figure~\ref{Fig6}. Each group represents H\,{\small II} regions at different evolutionary stages of star formation, and their distribution in NGC\,628 is shown in Figure~\ref{Fig7} with the same color as the rectangles in Figure~\ref{Fig6}.

The top cyan dashed rectangle in Figure~\ref{Fig6} encloses s subsample of H\,{\small II} regions with $E(B-V)_{\rm IRX} > E(B-V)_{{\rm H}\alpha/{\rm H}\beta}$. We can see from Figure~\ref{Fig7} that noticeable offsets are found for these targets between their  ${\rm H}\alpha$  and 21\,$\mu$m peaks. Figure~\ref{Fig8} shows the histogram of central offsets between PHANGS-MUSE H$\alpha$ and PHANGS-JWST/MIRI 21\,$\mu$m. We suggest that these H\,{\small II} regions are in the early evolutionary stage of star formation, during which the newly-formed young massive stars are still largely enveloped by birth clouds. The ${\rm H}\alpha$ and ${\rm H}\beta$ emission from the surrounding gas ionized by the UV ionizing photons hardly penetrates the dense clouds, and only a little fraction of the UV photons can escape from some holes of the star-forming clouds, leading to relatively low $E(B-V)_{{\rm H}\alpha/{\rm H}\beta}$ and high $E(B-V)_{\rm IRX}$. \citet{Chevance2020} found the typical lifetimes of molecular clouds are 10--30\,Myr, and the phase without H$\alpha$ emission is about 75--90 per\,cent of the cloud lifetime.

On the other hand, H\,{\small II} regions in the middle purple solid rectangle in Figure~\ref{Fig6} exhibit $E(B-V)_{\rm IRX}$ and $E(B-V)_{{\rm H}\alpha/{\rm H}\beta}$ following the relation as observed among local SFGs, adhering to the $R_{\rm EBV} = 0.51$  \citep{Qin2019b}. We can see from Figure~\ref{Fig7} that the peaks of the ${\rm H}\alpha$ emission from these H\,{\small II} regions largely overlap with the peaks at 21\,$\mu$m. And the central offsets of the majority of the HII regions concentrate at less than 40\,pc, shown in Figure~\ref{Fig8}. We consider that these H\,{\small II} regions are in the middle stage of star formation, lasting 1--5\,Myr \citep{Chevance2020}. During the on-going process of star formation, stellar winds drive  part of the birth clouds away, allowing us to observe both Balmer line emission, and the IR and UV radiation. As a result, the properties exhibited by these H\,{\small II} regions are generally consistent with the overall dust attenuation characteristics of the local SFGs.  

For the third class of H\,{\small II} regions as enclosed in the bottom red dashed-dotted rectangle in Figure~\ref{Fig6}, their $E(B-V)_{\rm IRX}$ and $E(B-V)_{{\rm H}\alpha/{\rm H}\beta}$ are both relatively low. These H\,{\small II} regions are often distributed slightly away from the bridges of spiral arms of NGC\,628, and most of them do not have bright 21\,$\mu$m dust emission, distinct from the first two classes of H\,{\small II} regions. This might indicate that these H\,{\small II} regions are in the late evolutionary stage of star formation, during which nearly all of the birth clouds have been blown away by stellar winds, leaving the young stars least obscured.  The Balmer line emission and dust IR emission become much weaker, leading to smaller values for $E(B-V)_{\rm IRX}$ and $E(B-V)_{{\rm H}\alpha/{\rm H}\beta}$.

\begin{figure*}
   \centering
   \includegraphics[width=0.8\textwidth]{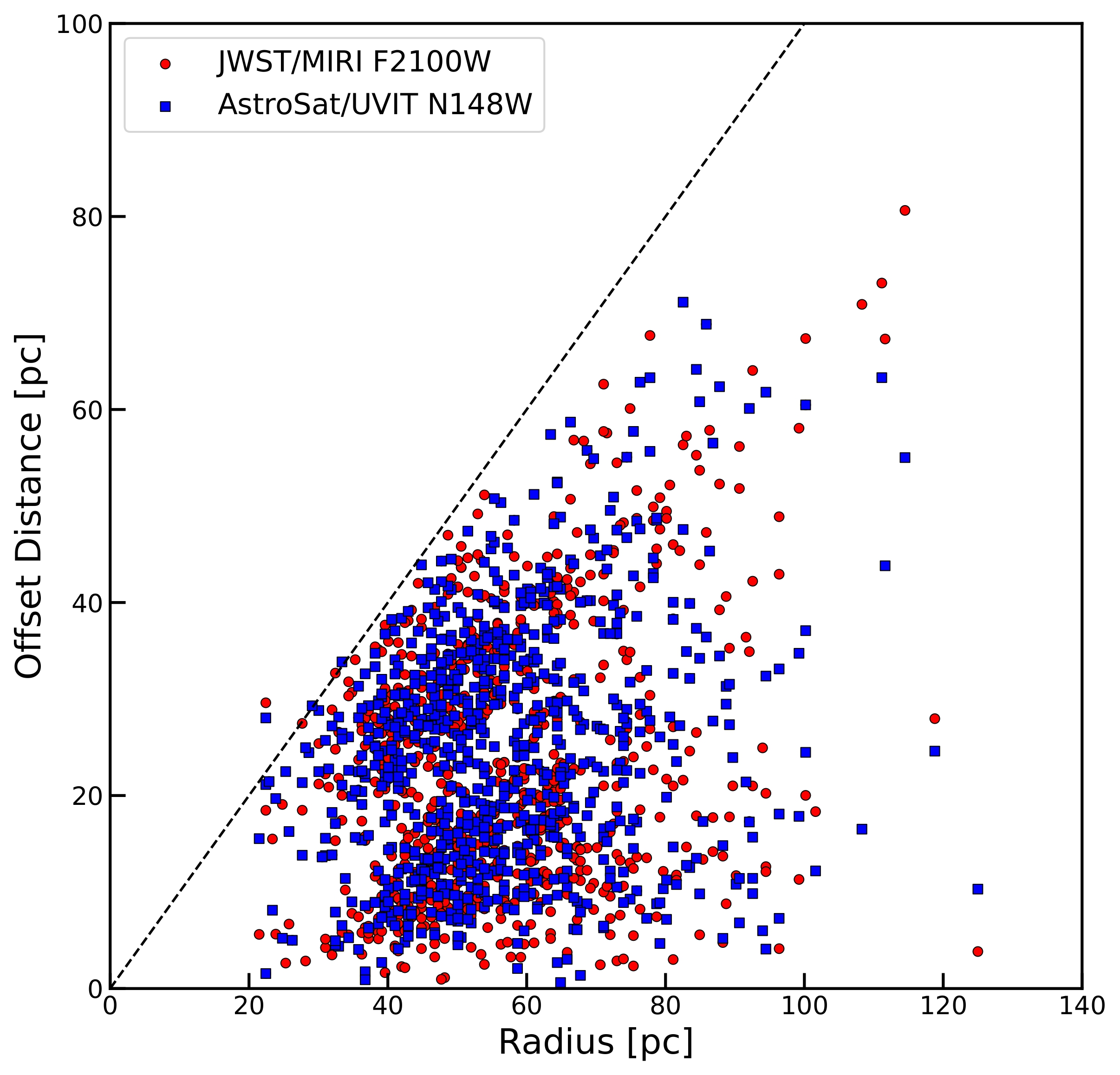}
   \caption{The central offsets of H\,{\small II} regions between ionized gas and stellar/dust components. The red circles represent the central offsets between PHANGS-MUSE H$\alpha$ and PHANGS-JWST/MIRI F2100W band, while the blue squares represent those between PHANGS-MUSE H$\alpha$ and PHANGS-AstroSat/UVIT N148W band. The dashed black line marks the 1:1 line.}
   \label{Fig9}
\end{figure*}

However, the positions of these H\,{\small II} regions in UV (young stars) and IR (dust-reprocessed light) images may differ due to wavelength-dependent resolution, dust obscuration, or physical offsets between ionized gas and stellar/dust components. We calculate the central offsets of each H\,{\small II} region between PHANGS-MUSE H$\alpha$ and PHANGS-JWST/MIRI F2100W band, as well as those between PHANGS-MUSE H$\alpha$ and PHANGS-AstroSat/UVIT N148W band, as shown in Figure~\ref{Fig9}. We find that the central offsets of H\,{\small II} regions between different bands exhibit a random distribution, with no statistically significant correlation to the physical sizes of the H\,{\small II} regions.

\section{Discussion}
\label{sect:discussion}

\begin{figure*}
   \centering
   \includegraphics[width=0.8\textwidth]{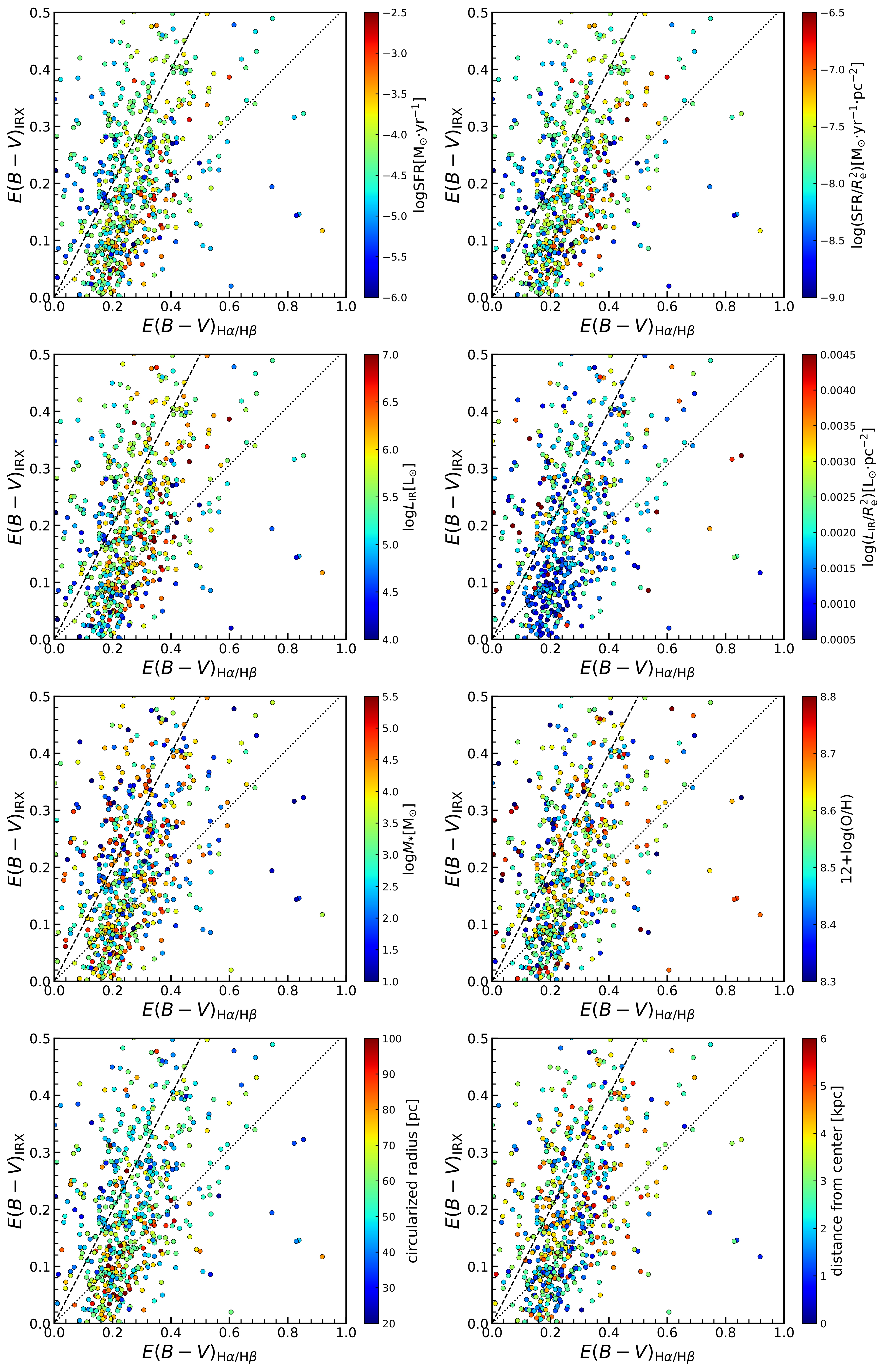}
   \caption{Comparison between $E(B-V)_{\rm IRX}$ and $E(B-V)_{{\rm H}\alpha/{\rm H}\beta}$ for a sample of 609 H\,{\small II} regions in NGC\,628. In these panels, from left to right and top to bottom, the parameters used for color coding are SFR, SFR surface density, $L_{\rm IR}$, $L_{\rm IR}$ surface density, stellar mass, gas-phase metallicity, radius, and the distance to galactic center, respectively. The dashed and dotted lines mark the relation of $E(B-V)_{\rm IRX} = E(B-V)_{{\rm H}\alpha/{\rm H}\beta}$ (i.e., $R_{\rm EBV} = 1$) and $E(B-V)_{\rm IRX} = 0.51 \times E(B-V)_{{\rm H}\alpha/{\rm H}\beta}$ (i.e., $R_{\rm EBV} = 0.51$).}
   \label{Fig10}
\end{figure*}

We investigate the relationships between two indicators IRX ($E(B-V)_{\rm IRX}$) and Balmer decrement ($E(B-V)_{{\rm H}\alpha/{\rm H}\beta}$) for 609 H\,{\small II} regions at scales of $\sim 50-200$\,pc in NGC\,628. We find that the relation $E(B-V)_{\rm IRX} = 0.51\,E(B-V)_{{\rm H}\alpha/{\rm H}\beta}$, which is established among local SFGs, does not hold for the H\,{\small II} regions, as illustrated in Figure~\ref{Fig6}. The scatter of data points in $E(B-V)_{\rm IRX}-E(B-V)_{{\rm H}\alpha/{\rm H}\beta}$ diagram is quite large, indicating that the two indicators break in probing dust attenuation  at spatial scales of H\,{\small II} regions with diameters of $50-200$\,pc in NGC\,628.  For heavily obscured H\,{\small II} regions in the early evolutionary stage, IRX is a good probe for dust attenuation but  ${\rm H}\alpha/{\rm H}\beta$ no longer works because these H\,{\small II} regions are optically thick.  As a result, the data points of our sample  H\,{\small II} regions do not follow a tight correlation between $E(B-V)_{\rm IRX}$ and $E(B-V)_{{\rm H}\alpha/{\rm H}\beta}$. When counting dust attenuation over regions of $> \sim 0.5$\,kpc scales, the integrated dust attenuation traced by  IRX and ${\rm H}\alpha/{\rm H}\beta$ converges to the global correlation among the local SFGs.  This confirms that the correlation between $E(B-V)_{\rm IRX}$ and $E(B-V)_{{\rm H}\alpha/{\rm H}\beta}$  holds at scales of $> \sim 0.5$\,kpc and breaks at scales of $\sim 50-200$\,pc in NGC\,628. Note that we do not attempt to determine a critical scale below which the correlation starts to deviate from the regime occupied by the local SFGs, because both the local SFGs and our sample H\,{\small II} regions show large spreads in size. 

Interestingly, \citet{Duffy2023} found that the spatially-resolved kpc-sized star-forming regions in nearby galaxies follows nearly the same IRX-$\beta$ relation as that for SFGs, indicating that the dust attenuation curve for the integrated light of galaxies  remains unchanged for the kpc-scale star-forming regions (see also \citealt{Ye2016}). This is consistent with our results.

Furthermore, we address the connections of dust attenuation with other parameters for our 609 H\,{\small II} regions in NGC\,628.  This is done by color coding the data points in the diagram of  $E(B-V)_{{\rm H}\alpha/{\rm H}\beta}$  versus $E(B-V)_{\rm IRX}$  by different physical parameters, including SFR, SFR surface density, $L_{\rm IR}$, $L_{\rm IR}$ surface density, stellar mass, gas-phase metallicity, radius, and the distance to galactic center for each H\,{\small II} region. The results are shown in Figure~\ref{Fig10}. Overall, we  find no apparent color pattern in the six panel, meaning that there is no statistically-significant  dependence of the ratio of $E(B-V)_{\rm IRX}$ to $E(B-V)_{{\rm H}\alpha/{\rm H}\beta}$ on these parameters. These results suggest that the discrepancy between IRX and ${\rm H}\alpha/{\rm H}\beta$  of H\,{\small II} regions,  i.e. the location in the $E(B-V)_{\rm IRX}$ versus $E(B-V)_{{\rm H}\alpha/{\rm H}\beta}$ diagram, is not controlled or influenced by these physical parameters. In contrast, the ratio exhibits a clear dependence on the sSFR surface density over galactic scales, as reported in \citet{Qin2019b}. This dependence no longer holds for individual H\,{\small II} regions at scales of $\sim 50-200$\,pc in NGC\,628. For a galaxy, it includes not only young stars but also intermediate-age and old stars, star-forming regions, and diffuse ISM. Therefore, the overall average ratio between $E(B-V)_{\rm IRX}$ and $E(B-V)_{{\rm H}\alpha/{\rm H}\beta}$ depends on the sSFR surface density. However, when focusing on individual star-forming regions, they only contain young stars and birth clouds at different evolutionary stages, possibly exhibiting distinct IRX and H$\alpha$/H$\beta$ from these of the entire galaxy. As a  consequence,  the correlation between $E(B-V)_{\rm IRX}$ and $E(B-V)_{{\rm H}\alpha/{\rm H}\beta}$ breaks at scales of individual star-forming regions.

In the early stage of star formation, young massive stars remain embedded within dense natal clouds (optically thick). This leads to high $E(B-V)_{\rm IRX}$ but low $E(B-V)_{{\rm H}\alpha/{\rm H}\beta}$ due to limited H$\alpha$ and H$\beta$ photons escape. In contrast, in the late stage of star formation, the surrounding dust is cleared via stellar winds \citep[e.g.][]{Chevance2020}, leading to lower $E(B-V)_{\rm IRX}$ and $E(B-V)_{{\rm H}\alpha/{\rm H}\beta}$ as IR, H$\alpha$, H$\beta$ emissions decrease. This evolutionary sequence explains the observed scatter in Figure~\ref{Fig6}. 
At H\,{\small II} region scales, the homogeneous dust-star mixture assumed in galaxy-scale studies breaks down. Clumpy dust distributions \citep{Witt2000} may decouple the attenuation of UV photons (traced by IRX) from the ionized gas (traced by H$\alpha$/H$\beta$). 
Stellar winds can create cavities in the ISM, altering the relative distribution of ionized gas and dust. H$\alpha$ emission primarily traces low-density ionized gas outside dense clouds, whereas IRX remains sensitive to embedded dust within optically thick regions. This spatial decoupling could contribute to the non-linear $E(B-V)_{\rm IRX}-E(B-V)_{{\rm H}\alpha/{\rm H}\beta}$ relation.

\section{Summary}
\label{sect:summary}

We use the UV, NIR and MIR imaging data from PHANGS-AstroSat and PHANGS-JWST, as well as the ionized nebula catalogue derived from PHANGS-MUSE, to spatially resolve 609 H\,{\small II} regions  in NGC\,628 and examine the connections between  IRX ($E(B-V)_{\rm IRX}$) and Balmer decrement ($E(B-V)_{{\rm H}\alpha/{\rm H}\beta}$) in tracing dust attenuation. Our main results are summarized as follows:

\begin{enumerate}

	\item The sample H\,{\small II} regions widely spread in the $E(B-V)_{\rm IRX}$ versus $E(B-V)_{{\rm H}\alpha/{\rm H}\beta}$ diagram,  indicating that the relation between IRX ($E(B-V)_{\rm IRX}$) and Balmer decrement ($E(B-V)_{{\rm H}\alpha/{\rm H}\beta}$) (i.e., $E(B-V)_{\rm IRX} = 0.51\,E(B-V)_{{\rm H}\alpha/{\rm H}\beta}$) on the galaxy scale of local SFGs no longer holds for H\,{\small II} regions at scales of $\sim 50-200$\,pc.
	
	\item  No dependence is found for the ratio of $E(B-V)_{\rm IRX}$ to $E(B-V)_{{\rm H}\alpha/{\rm H}\beta}$ on physical parameters SFR, SFR surface density, IR luminosity, IR luminosity surface density, stellar mass, gas-phase metallicity, circularized radius, and distance to galactic center for the sample H\,{\small II} regions. And we find no correlations between these parameters and IRX or ${\rm H}\alpha/{\rm H}\beta$.
	
	\item We point out that the H\,{\small II} regions with $E(B-V)_{\rm IRX}$ higher than $E(B-V)_{{\rm H}\alpha/{\rm H}\beta}$ are likely in early evolutionary stage in which the star-forming regions are heavily obscured and optically thick, while about three fifths of the regions are in the parameter range that local SFGs spread.  Note that a small portion of the H\,{\small II} regions exhibit very small ratio of $E(B-V)_{\rm IRX}$ to $E(B-V)_{{\rm H}\alpha/{\rm H}\beta}$, and we believe that these H\,{\small II} regions are in the late evolutionary stage in which the surrounding ISM are nearly all driven away by stellar winds. 

\end{enumerate}

\begin{acknowledgements}
   
   This work is supported by by the National Key Research and Development Program of China (2023YFA1608100),  the National Science Foundation of China (12233005, 12073078, 12173088, 12303015),  D.D.S. acknowledges the supports from Scientific Research Foundation for High-level Talents of Anhui University of Science and Technology (2024yjrc104), the National Science Foundation of Jiangsu Province (BK20231106) and the science research grants from the China Manned Space Project.  X.Z.Z. is grateful to the support from STCSM through grant No. 24DX1400100. 

\end{acknowledgements}

\bibliographystyle{raa}
\bibliography{bibtex}

\label{lastpage}

\end{document}